\documentclass[preprint,showpacs,preprintnumbers,amsmath,amssymb]{revtex4}

\usepackage{graphicx}% Include figure files
\usepackage{dcolumn}% Align table columns on decimal point
\usepackage{bm}% bold math

\begin{document}

\preprint{}
{\bf }\title{Relativistic predictions of 
exclusive $^{208}$Pb($\vec{p},2p$)$^{207}$T$\ell$ analyzing powers
at an incident energy of 202 MeV}
\author{G. C. Hillhouse$^{1,2}$, J. Mano$^{3}$, A. A. Cowley$^{1}$
and R. Neveling$^{1}$\footnote{\uppercase{P}resent address: 
\uppercase{T}homas \uppercase{J}efferson \uppercase{N}ational 
\uppercase{A}ccelerator \uppercase{F}acility, 
\uppercase{N}ewport \uppercase{N}ews,
\uppercase{V}irginia 23606, USA.} }  
\affiliation{$^{1}$Department of Physics, University 
of Stellenbosch, Private Bag X1, Matieland 7602, South Africa\\
$^{2}$Research Center for Nuclear Physics, Osaka
University, Ibaraki, Osaka 567-0047, Japan\\
$^{3}$Department of Electrical Engineering and 
Computer Science, Osaka Prefectural College of 
Technology, Osaka 572-8572, Japan}
\date{\today}

\begin{abstract}
Within the framework of the relativistic distorted wave 
impulse approximation (DWIA), we investigate the sensitivity of the analyzing power -  
for exclusive proton knockout from the 3s$_{1/2}$, 2d$_{3/2}$ and  2d$_{5/2}$ 
states in $^{208}$Pb, at an incident laboratory kinetic energy of 202 MeV, and for 
coincident coplanar scattering angles ($28.0^{\circ}$, $-54.6^{\circ}$) - to different 
distorting optical potentials, finite-range (FR) versus zero-range (ZR) approximations to the DWIA, as 
well as medium-modified coupling constants and meson masses. Results are also compared 
to the nonrelativistic DWIA predictions based on the Schr\"{o}dinger equation. Whereas the 
nonrelativistic model fails severely, both ZR and FR relativistic DWIA models provide an excellent 
description of the data. For the FR predictions, it is necessary to 
invoke a 20\% reduction of sigma-nucleon and omega-nucleon coupling constants  
as well as for $\sigma$-, $\rho$- and $\omega$-meson masses, by the nuclear medium. On the other hand,
the ZR predictions suggest that the strong interaction in the nuclear medium is adequately 
represented by the free nucleon-nucleon interaction associated with the impulse approximation.
We also demonstrate that, although the analyzing power is relatively insensitive to the use 
different relativistic global optical potential parameter sets, the prominent oscillatory 
behavior of this observable is largely attributed to distortion of the scattering wave 
functions relative to their plane wave values.
\end{abstract}


\maketitle

\section{\label{sec:intro}Introduction}
Recently, Neveling {\it et al.} \cite{Ne02} reported that both relativistic 
(Dirac equation) and nonrelativistic (Schr\"{o}dinger equation)
models, based on the distorted wave impulse approximation (DWIA), 
severely fail to reproduce exclusive ($\vec{p},2p$) analyzing power 
data for proton knockout from the 3s$_{1/2}$ and 2d$_{3/2}$  
states in $^{208}$Pb, at an incident laboratory kinetic energy of 202
MeV, and for coincident coplanar scattering angles ($28.0^{\circ}$, $-54.6^{\circ}$).  
For the prediction of energy-sharing cross sections, on the other hand, 
both dynamical models yield spectroscopic factors which are in 
good agreement with those extracted from ($e,e'p$) studies. 

Systematic corrections to the nonrelativistic model - such as 
different kinematic prescriptions for the nucleon-nucleon (NN) 
amplitudes, non-local corrections to the scattering wave functions, 
density-dependent modifications to the free NN scattering amplitudes, 
as well as the influence of different scattering and boundstate potentials - 
fail to remedy the analyzing power dilemma, and hence, it is 
not clear how to improve existing Schr\"{o}dinger-based analyses. 
However, such an exhaustive analysis has not yet been 
performed within the context of the relativistic DWIA and, hence, 
improvements to relativistic models could still prove to be important in
resolving the problem. Capitalizing on the fact that spin is an intrinsically relativistic 
phenomenon, as well as the success of Dirac phenomenology in describing the properties of 
nuclear matter, nuclear structure \cite{Se86}, as well as proton-induced 
spin observables for elastic- \cite{Mu87} and inelastic \cite{Ho88} scattering,
we focus in this paper, on systematic corrections to the DWIA based on the Dirac 
equation as the underlying dynamical equation of motion. Another advantage
of considering a relativistic approach is that both real and imaginary components of
the spin-orbit potential, which are crucial for describing analyzing powers
for s-state knockout in ($p,2p$) reactions, are directly related to the
Lorentz properties of mesons propagating the strong interaction. This microscopic
connection does not exist within the framework of the nonrelativistic Schr\"{o}dinger 
equation, where the spin-orbit interaction is usually introduced and adjusted merely to 
provide a good phenomenological description of elastic scattering data.

Motivated by the above considerations, we adopt a relativistic framework 
and study the sensitivity of exclusive analyzing powers to distorting 
optical potentials, finite-range (FR) versus zero-range (ZR) approximations 
to the DWIA, as well as nuclear medium modifications to the NN 
interaction. As already mentioned, we focus specifically on proton knockout from
the 3s$_{1/2}$, 2d$_{3/2}$ and 2d$_{5/2}$ states in $^{208}$Pb 
at an incident laboratory kinetic energy of 202 MeV, and for coincident 
coplanar scattering angles ($28.0^{\circ}$, $-54.6^{\circ}$).
Predictions are naturally compared to the corresponding nonrelativistic results.

One of the most challenging problems in nuclear physics is to 
understand how the properties of the strong interaction 
are modified inside nuclear matter. Various 
theoretical models \cite{Se86,Br91,Fu92} predict the modification 
of coupling constants as well as nucleon and meson masses
in normal nuclear matter. To date there is no direct
experimental evidence supporting these predictions. However, 
the exclusive nature of $(p,2p)$ reactions can be exploited 
to knockout protons from deep- to low-lying single particle 
states in nuclei, thus yielding information on the density 
dependence of the NN interaction \cite{Ha97}, and hence providing a 
stringent testing ground for theoretical models. 
In order to extract reliable information on NN 
medium modifications, it is important to understand the role 
of various approximations and model ingredients of the DWIA. 
In particular, an explanation for the failure of the
relativistic predictions reported in Ref. \cite{Ne02}, could 
possibly be attributed to the use of unreliable relativistic microscopic 
optical potentials for generating the scattering wave functions 
of the Dirac equation (see Sec. \ref{sec:rdwia}). To assess
the validity of this conjecture, we study the sensitivity of the 
analyzing power to different relativistic distorting potentials. 
Furthermore, current qualitative arguments suggest that, since 
analyzing powers are ratios of polarized cross sections, 
distortion effects on the scattering wave functions 
effectively cancel, and hence simple plane wave models 
(ignoring nuclear distortion) should be appropriate for 
studying polarization phenomena \cite{Ho88,Ho94}. 
This claim, however, has never been studied quantitatively, 
within the context of relativistic models and, hence we study this 
issue by comparing the distorted wave results 
of the analyzing power to corresponding plane 
wave predictions for zero scattering potentials. Our choice of a heavy 
target nucleus, $^{208}$Pb, and a relatively low incident energy of 202 MeV, is 
ideally suited for maximizing the influence of distortion effects, 
while still maintaining the validity of the impulse approximation, 
and also avoiding complications associated with the inclusion of 
recoil corrections in the relativistic Dirac equation \cite{Co93a,Ma93a}.

In principle, a FR approximation is more sophisticated than a ZR
approximation to the DWIA. However, in practice, FR predictions are subject 
to numerical errors due to extensive computational 
procedures (compared to ZR calculations). In this paper, we study the 
sensitivity of the analyzing power to both FR and ZR approximations. 
The only existing study in this regard was done by Ikebata \cite{Ik95} 
for the knockout of 1d$_{3/2}$ and 1d$_{5/2}$ protons from $^{40}$Ca at 
incident energies of 200 MeV and 300 MeV; no definite conclusion could be drawn 
as to which model gives a consistently better description of the data. For the case 
of a nucleus with a larger radius, such as $^{208}$Pb, we expect more 
pronounced differences between ZR and FR predictions. For estimating 
the influence of nuclear medium modifications of the NN interaction on
the analyzing power, we adopt the Brown-Rho scaling conjecture \cite{Br91} 
which attributes nuclear-medium modifications of coupling constants, 
as well as nucleon- and meson-masses, to partial restoration of chiral 
symmetry. An additional aim of this paper is to identify whether the 
analyzing power is an observable which demonstrates a preference
for the Schr\"{o}dinger- or Dirac equation as the underlying
dynamical equation.

\section{\label{sec:rdwia}
Relativistic Distorted Wave Impulse Approximation}
Both ZR and FR approximations to the relativistic DWIA have been
discussed in detail in Refs. \cite{Ik95} and \cite{Ma98}, respectively. 
We briefly describe the main ingredients of these models. For notational 
purposes, we denote an exclusive $(p,2p)$ reaction by $A(a,a'b)C$, whereby 
an incident proton ($a$) knocks out a bound proton ($b$)
from a specific orbital in the target nucleus ($A$) resulting in 
three particles in the final state, namely the recoil residual 
nucleus ($C$) and two outgoing protons, $a'$ and $b$, which are 
detected in coincidence at coplanar laboratory scattering angles 
(on opposite sides of the incident beam), $\theta_{a'}$ and $\theta_{b}$, 
respectively. All kinematic quantities are completely determined by specifying 
the rest masses, $m_{i}$, of particles, where 
$i$ = ($a$, $A$, $a'$, $b$, $C$), the laboratory kinetic energy, 
$T_{a}$, of incident particle $a$, the laboratory kinetic energy, 
$T_{a'}$, of scattered particle $a'$, the laboratory scattering angles 
$\theta_{a'}$ and $\theta_{b}$, and also the binding energy 
of the proton that is to be knocked out of the target nucleus.

For a FR approximation to the DWIA, the relativistic distorted wave 
transition matrix element is given by
\begin{eqnarray}
\hspace{-7mm} T_{L J M_J}(s_{a}, s_{a'}, s_{b}) & = & 
\int d\vec{r}\ d\vec{r}\,'\  
[
\bar{\psi}^{(-)}(\vec{r}, \vec{k}_{a'C},s_{a'}) 
\otimes   
\bar{\psi}^{(-)}(\vec{r}\,', \vec{k}_{bC},s_{b})\,
]\,  \times \nonumber\\
 & & \ \ \ \hat{t}_{NN}(\, | \vec{r} - \vec{r}\,'| \, )
[
\psi^{(+)}(\vec{r}, \vec{k}_{aA}, s_{a})
\otimes
\phi^{B}_{L J M_J}(\vec{r}\,'\,)
]\, ,
\label{e-frtjlm}
\end{eqnarray}
where $\otimes$ denotes the Kronecker product. The four-component
scattering wave functions, $\psi(\vec{r}, \vec{k}_{i},s_{i})$, 
are solutions to the fixed-energy Dirac equation with spherical 
scalar and time-like vector nuclear optical potentials: 
$\psi^{(+)}(\vec{r}, \vec{k}_{aA},s_{a})$ 
is the relativistic scattering  wave function of the incident
particle, $a$, with outgoing boundary conditions [indicated by the 
superscript $(+)$], where $\vec{k}_{aA}$ is the momentum of particle 
$a$ in the ($a$ + $A$) center-of-mass system, and $s_{a}$ is the 
spin projection of particle $a$ with respect to $\vec{k}_{aA}$ as the 
$\hat{z}$-quantization axis;  $\bar{\psi}^{(-)}(\vec{r}, \vec{k}_{j C},s_{j})$
 is the adjoint relativistic scattering wave function for particle $j$ 
[ $j$ = ($a',b$)] with incoming boundary conditions [indicated by the 
superscript $(-)$], where $\vec{k}_{j C}$ is the momentum of particle
$j$ in the ($j$ + $C$) center-of-mass system, and $s_{j}$ is the spin projection 
of particle $j$ with respect to $\vec{k}_{j C}$ as the $\hat{z}$-quantization axis.
The boundstate proton wave function, $\phi^{B}_{L J M_J}(\vec{r}\, )$, 
with single-particle quantum numbers $L$, $J$, and $M_{J}$, is obtained 
via selfconsistent solution to the Dirac-Hartree field equations of 
quantum hadrodynamics \cite{Ho81}. In addition, we adopt the impulse
approximation which assumes that the form of the NN scattering matrix
in the nuclear medium is the same as that for free NN scattering: the 
antisymmetrized NN scattering matrix, $\hat{t}_{NN}(\, | \vec{r} - \vec{r}\,'| \, )$, 
is parameterized in terms of five Lorentz invariants (scalar, pseudoscalar, 
vector, axial-vector, tensor). In principle, the NN $t$-matrix can be 
obtained via solution of the Bethe-Salpeter equation, where the on-shell 
NN amplitudes are matrix elements of this t--matrix. However, the complexity 
of this approach gives limited physical insight into the resulting amplitudes.
An alternative approach is to fit the amplitudes directly with some 
phenomenological form, rather than generating the $t$-matrix from a 
microscopic interaction. Although the microscopic approach is certainly 
more fundamental, the advantage of phenomenological fits lies in their 
simple analytical form, which allows them to be conveniently incorporated 
in calculations requiring the NN t-matrix as input. The NN $t$-matrix 
employed in this paper is based on the relativistic meson-exchange model 
described in Ref. \cite{Ho85}, the so-called relativistic Horowitz-Love-Franey 
(HLF) model, whereby the direct and exchange contributions to the amplitudes 
are parameterized separately in terms of a number of Yukawa-type meson exchanges 
in first-order Born approximation. The parameters of this interaction, namely 
the meson masses, meson-nucleon coupling constants and the cutoff parameters, 
have been adjusted to reproduce the free NN elastic scattering observables.

Adopting a much simpler ZR approximation, namely
\begin{eqnarray}
\hat{t}_{NN}(\, | \vec{r} - \vec{r}\,'| \, )\ =\ 
\hat{t}_{NN}(T_{\rm eff}^{\rm \ell ab}, \theta_{\rm eff}^{\rm cm})
\, \delta(\vec{r} - \vec{r}\,'\, )\, ,
\label{e-zra}
\end{eqnarray} 
the relativistic distorted wave transition matrix element in Eq.~(\ref{e-frtjlm})
reduces to
\begin{eqnarray}
\hspace{-7mm} T_{L J M_J}(s_{a}, s_{a'}, s_{b}) & = & 
\int d\vec{r}\, 
[\,
\bar{\psi}^{(-)}(\vec{r}, \vec{k}_{a'C},s_{a'}) 
\otimes   
\bar{\psi}^{(-)}(\vec{r}, \vec{k}_{bC},s_{b})\,] \nonumber \\
 & & \hat{t}_{NN}(T_{\rm eff}^{\rm \ell ab}, \theta_{\rm eff}^{\rm cm})\, 
[\,
\psi^{(+)}(\vec{r}, \vec{k}_{aA}, s_{a})
\otimes
\phi^{B}_{L J M_J}(\vec{r}\,)
\, ]\, ,
\label{e-tjlm}
\end{eqnarray}
where $T_{\rm eff}^{\rm \ell ab}$ and $\theta_{\rm eff}^{\rm cm}$ 
represent the effective two-body laboratory kinetic energy and 
effective center-of-mass scattering angles, respectively. 

As already mentioned, a FR approximation to the DWIA is inherently more 
sophisticated than a ZR approximation. However, in practice, the numerical 
evaluation of the six-dimensional FR transition matrix elements, 
given by Eq.~(\ref{e-frtjlm}), is nontrivial and subject 
to numerical uncertainties. On the other hand, for the ZR approximation, 
the three-dimensional integrand given by Eq.~(\ref{e-tjlm}), 
ensures numerical stability and rapid convergence (and hence faster 
computational time). Another  advantage of the ZR approximation is that 
one can directly employ experimental NN scattering amplitudes, rather than rely on a 
relativistic meson-exchange model, and hence, one is insensitive to 
uncertainties associated with interpolations and/or extrapolations of the limited 
meson-exchange parameter sets. In this paper, we compare FR and ZR 
predictions of the analyzing power.

The scalar and vector scattering potentials employed in the relativistic 
FR-DWIA calculations reported in Ref. \cite{Ne02} are microscopic in the 
sense that they are generated by folding the NN t-matrix, based on the HLF 
model, with the appropriate Lorentz densities via the $t \rho$ approximation.
An attractive feature of the $t \rho$ approximation is selfconsistency, 
that is, the HLF model is used for generating both scattering amplitudes 
and optical potentials. However, for the kinematic region of interest to this 
paper, we consider it inappropriate to employ microscopic $t \rho$ optical potentials, 
the reason being that HLF parameter sets only exist at 135 MeV and 200 MeV, whereas 
optical potentials for the outgoing protons are required at energies ranging 
between 24 and 170 MeV. Thus, enforcing selfconsistency would involve large, and relatively
crude, interpolations/extrapolations, leading to inaccurate
predictions of the analyzing power, as evidenced in Ref. \cite{Ne02}. 
Furthermore, the validity of the impulse approximation,
to generate microscopic $t \rho$ optical potentials at 
energies lower than 100 MeV, is questionable. Hence, in this paper 
we consider only global Dirac optical potentials, as opposed 
to microscopic $t \rho$ optical potentials, for obtaining 
the scattering wavefunctions of the Dirac equation. 

For studying medium effects on the NN interaction,
we make use of the scaling relations proposed by Brown and Rho 
\cite{Br91}, and also applied by Krein {\it et al.} \cite{Kr95}
to ($p,2p$) reactions, namely
\begin{eqnarray}
\frac{m_{\sigma}^{\ast}}{m_{\sigma}}
& \approx &  
\frac{m_{\rho}^{\ast}}{m_{\rho}} \approx \frac{m_{\omega}^{\ast}}
{m_{\omega}}\equiv \xi \label{eqn:a2}\ ,\\
\frac{g_{\sigma N}^{\ast}}{g_{\sigma N}}
 & \approx &
\frac{g_{\omega N}^{\ast}}{g_{\omega N}} \equiv \chi \label{eqn:a3}\ ,
\end{eqnarray}
where the medium-modified and free meson masses
are denoted by $m_{i}^{\ast}$ and $m_{i}$, with
$i \in \mbox{(}\sigma,\rho,\omega\mbox{)}$, respectively. Meson-nucleon 
coupling constants, with and without nuclear medium modifications, are 
denoted by $g_{j N}^{\ast}$ and $g_{j N}$, where $j \in \mbox{(}\sigma,\omega\mbox{)}$, 
respectively. 

The spin observable of interest, the analyzing power ($A_{y}$), is defined as
\begin{eqnarray}
A_{y}\ =\ \frac{\mbox{Tr}(T\,  \sigma_{y}\, T^{\dagger})}
{\mbox{Tr}(T T^{\dagger})}\,,
\label{e-dipj}
\end{eqnarray}
where $\sigma_{y}$ is the usual Pauli matrix, and
the $2 \times 2$ matrix, $T$, is given by
\begin{eqnarray}
T\ =\ \left(
\begin{array}{cc}
T_{L J}^{s_{a} = +\frac{1}{2}, s_{a'} = +\frac{1}{2}} & 
T_{L J}^{s_{a} = -\frac{1}{2}, s_{a'} = +\frac{1}{2}} \\
T_{L J}^{s_{a} = +\frac{1}{2}, s_{a'} = -\frac{1}{2}} & 
T_{L J}^{s_{a} = -\frac{1}{2}, s_{a'} = -\frac{1}{2}} 
\end{array}\right)\, ,
\label{e-ttwobytwo}
\end{eqnarray}
where $s_{a} = \pm \frac{1}{2}$ and $s_{a'} = \pm \frac{1}{2}$ refer to the spin
projections of particles $a$ and $a'$ along the 
$\hat{k}_{a A}=\hat{z}$ and $\hat{k}_{a' C}=\hat{z}'$ 
quantization axes, respectively, and the $\hat{y}$-axis is defined by
$\hat{k}_{a A} \times\hat{k}_{a' C}$. The matrix elements, 
$T_{L J}^{s_{a}, s_{a'}}$, are related to the relativistic $(p,2p)$ 
ZR- and FR-DWIA transition matrix elements, 
$T_{L J M_{J}}(s_{a}, s_{a'}, s_{b})$, defined by 
Eqs.~(\ref{e-frtjlm}) and (\ref{e-tjlm}) respectively, via
\begin{eqnarray}
T_{L J}^{s_{a}, s_{a'}}\ =\ \sum_{M_J, s_b} T_{L J M_J}
(s_a, s_{a'}, s_b)\,.
\label{e-tljmrelatet}
\end{eqnarray}

\section{\label{sec:results}Results}
In this section we investigate the sensitivity of the analyzing power - 
for the knockout of protons from the 3s$_{1/2}$, 2d$_{3/2}$ 
and 2d$_{5/2}$ states in $^{208}$Pb, at an incident energy of 
202 MeV, and for coincident coplanar scattering angles 
($28.0^{\circ}$, $-54.6^{\circ}$) - to distorting optical 
potentials, FR versus ZR approximations to the relativistic 
DWIA, as well as to medium-modified coupling constants and 
meson masses. We also compare our relativistic results to 
nonrelativistic DWIA predictions. Unless otherwise 
specified, all DWIA predictions are based on the energy-dependent 
global Dirac optical potential parameter set which has been constrained 
by $^{208}$Pb($p,p$) elastic scattering data for incident proton 
energies between 21 MeV and 1040 MeV \cite{Co93b}.

\begin{figure}[htb]
\includegraphics{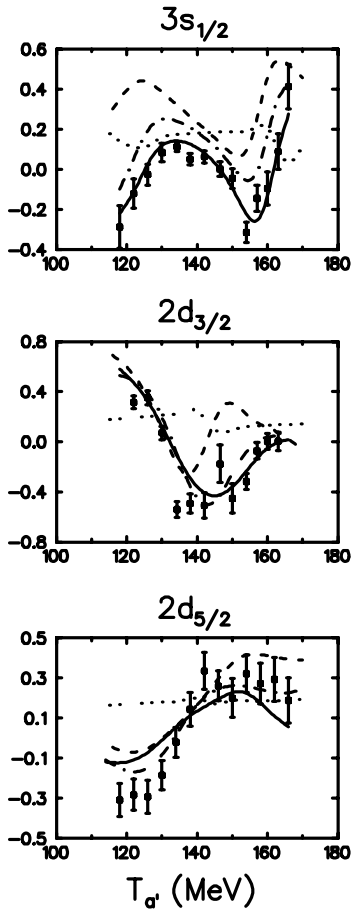}
\caption{\label{fig:gch-fig1} 
Analyzing powers plotted as a function of the kinetic 
energy, $T_{a'}$, for the knockout of protons from 
the 3s$_{1/2}$, 2d$_{3/2}$ and 2d$_{5/2}$ states 
in $^{208}$Pb, at an incident energy of 202 MeV, and 
for coincident coplanar scattering angles ($28.0^{\circ}$, $-54.6^{\circ}$).
The different line types represent the following calculations:
relativistic ZR-DWIA (solid line), relativistic plane wave (dotted
line), nonrelativistic DWIA (dashed line), and relativistic 
FR-DWIA (dot-dashed line): all calculations exclude 
medium-modified coupling constants and meson masses. 
The data are from Ref. \cite{Ne02}.}
\end{figure}

First, we display the influence of relativistic nuclear distortion 
effects by comparing relativistic ZR-DWIA 
predictions to corresponding plane wave predictions (with 
zero scattering potentials) for knockout from all three states: 
in Fig.~(\ref{fig:gch-fig1}), the solid line indicates the 
relativistic distorted wave result and the dotted line represents the 
relativistic plane wave result. We see that the prominent 
oscillatory structure of the analyzing powers is mostly 
attributed to distortions of the scattering wave functions.
This clearly illustrates the importance of nuclear distortion 
on the analyzing power, thus refuting, for the first time, qualitative 
claims that spin observables, being ratios of cross sections, are 
insensitive to nuclear distortion effects. In addition, we have
also investigated the sensitivity of the analyzing powers to 
a variety of different global Dirac optical potential parameter 
sets \cite{Co93b}. Although these results are not displayed, we 
found that the analyzing powers are relatively insensitive to 
different global optical potentials, with differences between 
parameter sets being smaller than the experimental statistical error.
For the reaction kinematics of interest, NN amplitudes need to be evaluated at
$T_{\rm eff}^{\rm \ell ab} \approx 180\ \mbox{MeV}$ and $\theta_{\rm eff}^{\rm cm} 
\approx 60^{\circ}$, but the closest HLF parameter sets exist at 135 MeV and 200 MeV.
To improve the accuracy of our FR predictions, we have generated a new HLF 
parameter set at 180 MeV by fitting to the experimental NN amplitudes \cite{Ho85}. 
We have checked the validity of the HLF parameter set by comparing ZR 
calculations based on the HLF model to corresponding calculations based 
directly on the experimental amplitudes: the predicted ($p,2p$) analyzing 
powers are identical.
\begin{figure}[htb]
\includegraphics{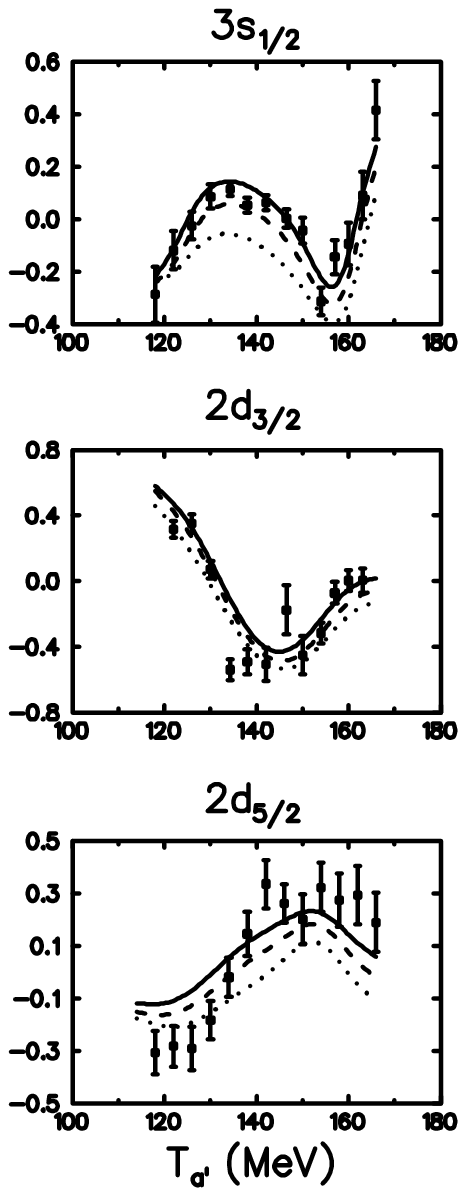}
\caption{\label{fig:gch-fig2} 
Analyzing powers plotted as a function of the kinetic 
energy, $T_{a'}$, for the knockout of protons from 
the 3s$_{1/2}$, 2d$_{3/2}$ and 
2d$_{5/2}$ states in $^{208}$Pb, at an incident energy of 202 MeV, and 
for coincident coplanar scattering angles ($28.0^{\circ}$, $-54.6^{\circ}$).
The different line types represent the following calculations:
relativistic ZR-DWIA excluding medium-modified coupling constants 
and meson masses (solid line), relativistic ZR-DWIA with a 10\%
reduction of the medium-modified coupling constants and meson masses 
(dashed line), and relativistic ZR-DWIA with a 20\%
reduction of the medium-modified coupling constants 
and meson masses (dotted line). 
The data are from Ref. \cite{Ne02}.}
\end{figure}
Next, we compare relativistic FR to relativistic ZR predictions, 
excluding medium modifications to the NN interaction. 
In Fig.~(\ref{fig:gch-fig1}), we see that the ZR prediction (solid line)
almost perfectly describes the data for knockout from the
3s$_{1/2}$ and 2d$_{3/2}$ states: recall that previous relativistic
and nonrelativistic models fail to reproduce these data \cite{Ne02}. 
For the 3s$_{1/2}$ state, the relativistic FR result (dot-dashed line) 
is consistently shifted above the data. 
Nevertheless, the relativistic FR prediction still provides a qualitative 
description of the data. For knockout from the 2d$_{3/2}$ and 2d$_{5/2}$ 
states both relativistic ZR and FR models describe the data reasonably well.

We also compare our relativistic calculations
to nonrelativistic  [dashed line in Fig.~(\ref{fig:gch-fig1})] 
DWIA predictions, excluding medium modifications 
of the NN interaction, recently 
reported in Ref. \cite{Ne02}: the nonrelativistic 
Schr\"{o}dinger-based calculations are based on 
the computer code THREEDEE by Chant and Roos \cite{Ch}.
With the exception of the 2d$_{5/2}$, it is clearly seen that the relativistic 
ZR (solid line) and FR (dot-dashed line) predictions in 
Fig.~(\ref{fig:gch-fig1}) are consistently superior compared 
to the corresponding nonrelativistic calculations. This 
suggests that the Dirac equation is the most appropriate 
dynamical equation for the description of analyzing powers. 
Moreover, these results represent the clearest signatures to date
for the evidence of relativistic dynamics 
in polarization phenomena.

\begin{figure}[htb]
\includegraphics{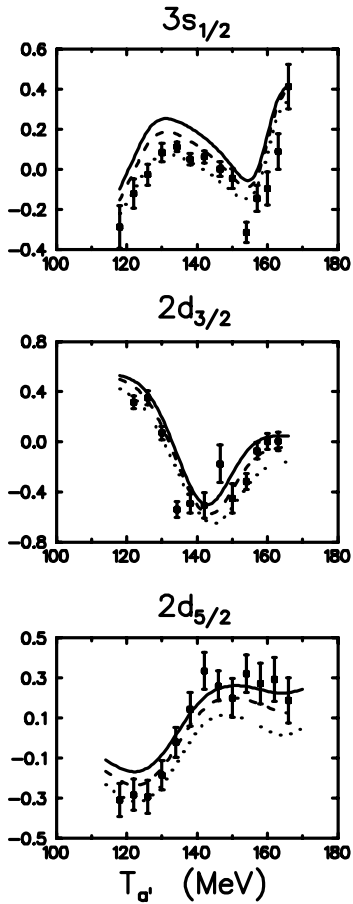}
\caption{\label{fig:gch-fig3} 
Analyzing powers plotted as a function of the kinetic 
energy, $T_{a'}$, for the knockout of protons from 
the 3s$_{1/2}$, 2d$_{3/2}$ and 
2d$_{5/2}$ states in $^{208}$Pb, at an incident energy of 202 MeV, and 
for coincident coplanar scattering angles ($28.0^{\circ}$, $-54.6^{\circ}$).
The different line types represent the following calculations:
relativistic FR-DWIA excluding medium-modified coupling constants 
and meson masses (solid line), relativistic FR-DWIA with a 10\%
reduction of the medium-modified coupling constants and meson masses 
(dashed line), and relativistic FR-DWIA with a 20\%
reduction of the medium-modified coupling constants 
and meson masses (dotted line). 
The data are from Ref. \cite{Ne02}.}
\end{figure}

Although the ($p,2p$) reaction of interest are mainly surface peaked,
radial localization (radial contribution of the reaction to DWIA cross section)
arguments \cite{Ne02} suggest that s-state knockout exhibits
a larger contribution from the nuclear interior than the d-states and,
hence, s-state knockout is more susceptible to nuclear
medium modifications of the NN interaction. Thus, the inclusion of
nuclear medium effects offers the possibility to improve the relativistic FR-DWIA 
prediction of the 3s$_{1/2}$ analyzing power.
We now study the sensitivity of the analyzing power to values of $\xi = \chi$ 
[see Eqs.~(\ref{eqn:a2}) and (\ref{eqn:a3})] less than unity for both relativistic ZR 
and relativistic FR approximations, that is, we assume that the effect of the nuclear medium
is to reduce values of the masses and coupling constants of certain 
mesons relative to their corresponding free values. Note that,
in principle the coupling constants and meson masses are independent
quantities and, hence there is no fundamental reason to set $\xi = \chi$.
The latter equality is only assumed for simplicity, so as to get a feeling
for the sensitivity of observables to changes in the relevant coupling 
constants and meson masses.
In Fig.~(\ref{fig:gch-fig2}), we display relativistic ZR-DWIA results for 
$\xi = \chi\ \in\ (0.9, 0.8)$ corresponding to reductions 
of the meson masses and coupling constants by 10\% (dashed line) and 20\% 
(dotted line), respectively: results excluding medium modifications are
indicated by the solid line.  The choice of values for $\xi$ and $\chi$ 
is motivated by the fact that the proton-knockout reactions of interest 
are mainly localized in the nuclear surface and, hence, the nuclear medium 
modifications are expected to play a relatively minor role. The corresponding 
relativistic FR-DWIA predictions are shown in Fig.~(\ref{fig:gch-fig3}). 
Although not displayed, we have already established that values of  
$\xi = \chi < 0.8$  fail to reproduce the analyzing powers for both 
ZR and FR approximations. For the FR calculations, we see that a reduction 
of meson masses and coupling constants by between 10\% (dashed line) to 
20\% (dotted line) consistently improves the predictions for knockout from 
all states: the agreement with the 3s$_{1/2}$ analyzing power is particularly 
impressive. Similar qualitative behavior was observed for the nonrelativistic distorted
wave predictions reported in Ref. \cite{Ne02}, where the inclusion of empirical
density-dependent correction to the analyzing power shifts predictions closer
to the data. In addition, by analyzing the ``effective polarization'',
for proton knockout from $^{16}$O and $^{40}$Ca at 200 MeV, within the framework 
of the nonrelativistic DWIA, Krein {\it et al.} \cite{Kr95} also reported 
similar evidence for the modification of meson masses and coupling constants
by the nuclear medium. On the other hand, relativistic ZR predictions without medium effects 
give a better description of the data: a 20\% reduction fails to
reproduce the 3s$_{1/2}$ analyzing power. In general, one can conclude that 
relativistic FR predictions with medium effects and relativistic ZR calculations excluding 
medium effects both give a satisfactory description of the data.
In order to make more definite statements on the importance of nuclear medium
effects for ($p,2p$) reactions, one needs to measure and interpret complete sets 
of spin observables, as opposed to only the analyzing power: this will be studied
in a future paper. Also, one needs to consider the knockout of protons from deeper
lying states in $^{208}$Pb, where the contribution from the nuclear interior
is more substantial.

\section{\label{sec:conclusions}Summary and conclusions}
 In this work we have focused on a relativistic distorted wave
description for exclusive proton knockout from the 
3s$_{1/2}$, 2d$_{3/2}$ and 2d$_{5/2}$ states in $^{208}$Pb, 
at an incident energy of 202 MeV, and for coincident 
coplanar scattering angles ($28.0^{\circ}$, $-54.6^{\circ}$). 
Previous relativistic and nonrelativistic models fail to describe
the analyzing power for 3s$_{1/2}$- and 2d$_{3/2}$-knockout \cite{Ne02}. 
Exhaustive corrections to the nonrelativistic model fail to resolve 
the dilemma. On the other hand, this is the first time that such a 
systematic analyses has now been performed within the context of 
the relativistic DWIA. We have identified two possible reasons 
for the failure of the relativistic FR-DWIA predictions 
reported in Ref. \cite{Ne02}. First of all, for the reaction kinematics 
of interest, the microscopic optical potentials generated via the $t \rho$ 
approach were not refined enough. Secondly, the influence of density-dependent
corrections to the NN interaction was previously not considered, and thus
previous relativistic FR-DWIA predictions \cite{Ne02} implicitly 
underestimated an important ingredient of the theoretical treatment.

These shortcomings have been addressed by employing appropriate global 
Dirac optical potentials and also by studying the role of 
medium-modified meson masses and 
coupling constants, constrained by the Brown-Rho scaling conjecture, for 
both relativistic ZR- and FR-DWIA calculations of the analyzing power. 
We also compare relativistic predictions to the nonrelativistic 
results quoted in Ref. \cite{Ne02}.

In this paper we have demonstrated  the superiority of the relativistic 
Dirac-equation, as compared to the nonrelativistic Schr\"{o}dinger equation, 
for the description of the exclusive ($\vec{p},2 p$) analyzing powers within
the context of the DWIA. Both relativistic ZR and FR approximations 
to the DWIA provide an excellent description of the
analyzing power data. On one hand, the relativistic ZR predictions suggest that the scattering 
matrix for NN scattering in the nuclear medium is adequately represented by 
the corresponding matrix for free NN scattering and, hence it is not necessary to consider
nuclear medium modifications to the NN interaction. On the other hand, the 
relativistic FR results suggest that a 10\% to 20\% reduction of meson-coupling constants 
and meson masses by the nuclear medium is essential for providing a consistent 
description of the 3s$_{1/2}$, 2d$_{3/2}$ and 2d$_{5/2}$ analyzing powers.
In order to extract more conclusive information regarding the influence
of the nuclear medium on the properties of the strong interaction, it is
necessary to study complete sets of polarization transfer observables
for the exclusive knockout of protons from deeper-lying states in a variety 
of nuclei.

We have also established that the analyzing power is relatively
insensitive to different global Dirac optical potential parameter sets. In
addition, by comparing relativistic ZR-DWIA predictions to corresponding
plane wave predictions (zero scattering potentials), we have demonstrated
the importance of distorting potentials for describing the oscillatory 
behavior of the analyzing powers, thus refuting qualitative arguments
that spin observables are insensitive to nuclear distortion effects.

\begin{acknowledgements}
G.C.H acknowledges financial support from the Japanese 
Ministry of Education, Science and Technology for research conducted 
at the Research Center for Nuclear Physics, Osaka University, Osaka, 
Japan. This material is based upon work supported by the National
Research Foundation under Grant numbers: GUN 2058507, 2053786, 2053773.
\end{acknowledgements}

\end{document}